\documentclass[aps,draft,twocolumn,eqsecnum,showpacs]{revtex4}
\input epsf
\newcommand{\be}{\begin{equation}}
\newcommand{\ee}{\end{equation}}
\newcommand{\bea}{\begin{eqnarray}}
\newcommand{\eea}{\end{eqnarray}}

\newcommand{\vk}{\vec{k}}
\newcommand{\vq}{\vec{q}}
\newcommand{\va}{\vec{a}}
\newcommand{\vf}{\vec{v}_F}
\newcommand{\vv}{\vec{v}}
\newcommand{\vrr}{\vec{r}}
\newcommand{\vn}{\hat{n}}
\newcommand{\vt}{\hat{t}}
\begin{document}
%%%%%%%%%%%%%%%%%%%%%%%%%%%%%%%%%%%%%%%%%%%%
\title{Strongly correlated fermions with nonlinear energy dispersion and spontaneous generation of anisotropic
phases}
%%%%%%%%%%%%%%%%%%%%%%%%%%%%%%%%%%%%%%%%%%%%%%%%%%%%%%%
\author{Daniel G.\ Barci}
\affiliation{Departamento de F\'\i sica Te\'orica,
Universidade do Estado do Rio de Janeiro,Rua S\~ao Francisco Xavier 524, 20550-
013, 
Rio de Janeiro, RJ, Brazil}
\author{Luis E.\ Oxman}
\affiliation{Departamento de F\'\i sica,
Universidade Federal Fluminense, Av. Litor\^anea  s/n, Boa Viagem 24214-340, Niter\'oi, RJ, Brazil}

\date{May 23, 2003}
%%%%%%%%%%%%%%%%%%%%%%%%%%%%%%%%%%%%%%%%%%%%%%%%%%%%%%%%%%%%%%%%%%%%%%%

\begin{abstract}
Using the bosonization approach we study fermionic systems with a nonlinear dispersion relation in dimension $d\ge 2$. We explicitly show how the band curvature gives rise to interaction terms in the bosonic version of the model. Although these terms are perturbatively irrelevant in relation to the Landau Fermi liquid fixed point, they become relevant perturbations when instabilities take place. Using a coherent state path integral technique we built up the effective 
action that governs the dynamics of the Fermi surface fluctuations. We consider the combined effect of fermionic interactions and band curvature on possible anisotropic phases triggered by negative Landau parameters (Pomeranchuck instabilities). In particular we study in some detail the phase diagram for the isotropic/nematic/hexatic quantum phase transition.
\end{abstract}


%71.10.Hf Non-Fermi-liquid ground states, electron phase diagrams and phase transitions in model %systems
%71.10.Pm Fermions in reduced dimensions (anyons, composite fermions, Luttinger liquid, etc.) %(for anyon mechanism in superconductors, see 74.20.Mf)
%05.30.Fk Fermion systems and electron gas (see also 71.10 Theories and models of many-electron %systems)
%71.27.+a Strongly correlated electron systems; heavy fermions

\maketitle

%%%%%%%%%%%%%%%%%%%%%%%%%%%%%%%%%%%%%%%%%%%%%%%%%%%%%%%%%%%%%%%%%%%%%%%

%%%%%%%%%%%%%%%%%%%%%%%%%%%%%%%%%%%%%%%%%%%%%%%%%%%%%%%%%%%%%%%%%%%%%%%
\section{Introduction}
%%%%%%%%%%%%%%%%%%%%%%%%%%%%%%%%%%%%%%%%%%%%%%%%%%%%%%%%%%%%%%%%%%%%%%%

One commonly used approximation to study strongly correlated fermions at 
low energies is the linearization of the fermion energy dispersion relation near the 
Fermi surface. For example, its implementation in the context of bosonization leads, together with the Renormalization Group (RG), to a powerful nonperturbative technique to deal with many-body problems. In one dimension this approximation gives rise to the Tomonaga-Luttinger model\cite{Haldane}, while in $d>1$ the Landau theory of Fermi liquids comes up as a fixed point in RG sense\cite{CNF1,CNF2,Marston,Marston2}. In both cases, the bosonized Hamiltonian is quadratic and the model can be solved 
exactly, while small perturbations can be studied using RG. In particular, nonlinear terms in the 
dispersion relation are perturbatively irrelevant, that is, they do not modify the long-wavelength properties of the system.

However, in some cases, the linear approximation must be improved. Induced nonlinear terms in the quasiparticle energy dispersion become important when fermions are coupled to transverse fluctuating gauge fields. This is the case of some models of  high $T_c$ superconductors\cite{HTc} and gauge theories of the half-filled ($\nu=1/2$) Quantum Hall Effect\cite{QHE1/2}.

More recently, it was suggested the possibility of having anisotropic ground states driven by spontaneous rotational symmetry breaking (Quantum Liquid Crystals)\cite{HTcliquidcrystal,QHliquidcrystal}. These new phases were proposed to describe transport properties of half-filled quantum Hall systems\cite{QHliquidcrystal} and  high $T_c$ superconductors\cite{HTcliquidcrystal}. They can be associated to Pomeranchuk instabilities\cite{Pomeranchuk} of the isotropic Fermi surface and for these novel ground states become stable a nonlinear fermion dispersion relation turns out to be essential\cite{NematicKFV}.
An interesting related phenomenon was also pointed out in ref. \cite{Metzner}, where a Hubbard model is studied using RG techniques. There, it was shown that for a certain region of the parameter space strong forward scattering interactions favor Pomeranchuk instabilities leading to a breakdown of the discrete rotational symmetry.

With these motivations in mind, we will present a systematic study of the bosonization of fermionic
systems with a nonlinear energy dispersion relation in any number of dimensions. 

Concerning the Luttinger model case ($d=1$), the first work were nonlinear dispersion terms were taken into account 
                 was carried out by Haldane\cite{Haldane}. He showed that the resulting bosonized Hamiltonian is modified by the addition of nonquadratic terms in the bosonic variables. Explicit corrections to the corresponding one-particle Green function were recently computed in ref.\ \cite{Kopietz1}. 
For dimensions greater than one, the influence of nonlinear dispersion terms on the one-particle Green function was 
studied by Kopietz in the framework of functional bosonization\cite{kopietz2}.

In this work we are interested in studying the dynamics of the Fermi surface in dimension $d\geq 2$ and getting an explicit understanding of how the nonlinear dispersion relation could stabilize possible phases, other than the isotropic Fermi liquid one. 

The idea of a Fermi surface as a dynamical quantum extended object was originally introduced by 
Luther\cite{Luther} and improved by Haldane\cite{HaldaneFermiSee1,HaldaneFermiSee2}. This concept was developed in great detail by Castro Neto and Fradkin\cite{CNF1,CNF2} and by Houghton and Marston\cite{Marston}. In ref.\ \cite{CNF1}, the bosonized theory is written in a coherent state basis $|\phi>$ representing deformations of the Fermi surface. In this way, the quantum dynamics of the system (the partition function) is expressed as a Feynman path integral where the ``sum over paths'' corresponds to summing up the contributions coming from all possible deformations of the Fermi surface. 

The first part of this paper is devoted to apply the abovementioned formalism to explicitly show how nonlinear terms in the energy dispersion relation contribute with interacting nonquadratic terms in the bosonized action, in arbitrary dimensions. 
Eq.\ \ref{lagrangianaqfinal} below is one of the main results of this paper, showing the effective low energy Lagrangian of the system. There, $\phi_q(\vk)$ represents deformations of the Fermi surface at the Fermi point $\vk$ (particle-hole excitations with momentum $\vq$). The second and third derivatives of the dispersion relation lead to the cubic and quartic bosonic terms, respectively. Notice that this general formulation stands for 
arbitrary smooth Fermi surfaces being particularly suitable for studying phases where shape deformations are present.

These deformations can be classified according to the symmetries of an order parameter, similarly to 
the classification of  classical liquid crystals\cite{deGennes}. In fact, we can think about the quantum equivalent of smectic\cite{smectic1,smectic2}, nematic\cite{nematic} or hexatic\cite{NematicKFV} phases and their corresponding quantum phase transitions.

In ref.\ \cite{NematicKFV}, V.\ Oganesian {\em et al.} showed for the first time that quantum isotropic/nematic and isotropic/hexatic phase transitions are possible in systems where the Landau parameters $F_2$ and $F_6$ of the usual Fermi liquid theory\cite{FL} assume large and negative values (for a definition of the $F_n$'s see Eq.\ \ref{Fn} below). In that reference, one important ingredient to stabilize the anisotropic states is the consideration of a nonlinear energy dispersion relation in the model Hamiltonian. The corresponding electronic properties are very promising since the quantum nematic and hexatic states seem to present non-Fermi liquid behavior\cite{NematicKFV}. 

For these reasons, in the second part of this paper we use the nonperturbative bosonization approach to study quantum phase transitions to anisotropic electronic states in two dimensional systems. In particular, we will concentrate in nematic and hexatic quantum liquid crystal phases where the order parameter is invariant under $\pi$ and $\pi/3$ rotations, respectively. 

The main result of this paper is displayed by the phase diagrams in figures \ref{bp} and \ref{tp}. By integrating out all the stable modes we obtain an effective free energy at zero temperature as a function of the Landau parameters $F_2$ and $F_6$. We find different behaviors 
depending on the relative values of the stable Landau parameters $F_n$ ($n\ne$ $2$, $6$). When these parameters are small, the phase diagram has a tricritical point where two second order phase transitions (isotropic/nematic, isotropic/hexatic) and a first order one (nematic/hexatic) meet together. 
However if the stable $F_n$'s are not small, a coexisting nematic-hexatic phase with a tetracritical point is possible. 

In the rest of the paper we explicitly develop the mathematical details leading to these results. 
In section \S \ref{Hamiltonian} we present our Hamiltonian model for spinless fermions. In section \S \ref{BosonizedHamiltonian} we show how to apply the bosonization method to Hamiltonians with a nonlinear dispersion relation and explicitly compute the corresponding nonquadratic bosonized terms. Then, in section \S \ref{Dynamics} we analyze the Fermi surface dynamics, building up a coherent-state path integral formulation for the partition function of the system. 
Finally, in section \S \ref{isotropic-nematic-hexatic} we analyze the possibility of isotropic-nematic-hexatic 
quantum phase transitions. Section \S \ref{Conclusions} is devoted to a discussion of the results and to the presentation of our conclusions.

%%%%%%%%%%%%%%%%%%%%%%%%%%%%%%%%%%%%%%%%%%%%%%%%%%%%%%
\section{The Hamiltonian}
\label{Hamiltonian}
%%%%%%%%%%%%%%%%%%%%%%%%%%%%%%%%%%%%%%%%%%%%%%%%%%%%%%
We consider a  fermionic system characterized by a smooth Fermi surface given by the set of Fermi points $\vk_F$ satisfying $\epsilon(\vk_F)=\mu$, where $\epsilon(\vk)$ is an arbitrary energy dispersion relation.
The one particle excitations are associated with a set of 
operators $c^{\dagger}_{\vk}$ and  $c_{\vk}$, creating and destroying a fermion with 
momentum $\vk$. These operators satisfy the usual fermionic anticommutation relations. 
For simplicity we ignore the spin degree of freedom, however the extension to spinfull fermions is straightforward.

In general, the Hamiltonian can be written in the form
\be
H=H_0+H_{\rm int}.
\label{H}
\ee
The free (quadratic) term is given by
\be
H_0=\sum_{\vk} (\epsilon(\vk)-\mu) c^{\dagger}_{\vk} c_{\vk}.
\label{H0}
\ee
A general two-body interaction term can be written as 
\be
H_{\rm int}=\frac{1}{2V}\sum_{\vk_F,\vk_F',\vq} f_{\vk_F,\vk_F'}(\vq) \;\;
c^{\dagger}_{\vk_F-\frac{\vq}{2}}c_{\vk_F+\frac{\vq}{2}}c^{\dagger}_{\vk_F'+\frac{\vq}{2}}c_{\vk_F'-\frac{\vq}{2}}, 
\label{Hint}
\ee
where $f_{\vk_F,\vk_F'}(\vq)$ is the scattering amplitude among two particle-hole pairs with momentum $\vq$, 
at the Fermi points $\vk_F$ and $\vk_F'$. 

In Eq.\ \ref{H0} the energy dispersion relation $\epsilon(\vk)$ can be expanded in powers  of
$\vq=\vk-\vk_F$, 
%\begin{widetext}
\bea
\epsilon(\vk)&=&\mu+\vf\cdot\vq+ 
\frac{1}{2}\left.\frac{\partial^2 \epsilon}{\partial k_i\partial k_j}\right|_{\vk=\vk_F}
\!\!\!\!\!\! q_i q_j  \nonumber \\
&+& \frac{1}{3!}\left.\frac{\partial^3 \epsilon}{\partial k_i\partial k_j\partial k_l}\right|_{\vk=\vk_F}
\!\!\!\!\!\! q_i q_j q_l+\ldots,
\label{dispersion}
\eea
%\end{widetext}
where $\vf=\vf (\vk_F)=\left.\vec{\nabla}\epsilon(\vk)\right|_{\vk=\vk_F}$ is the Fermi velocity.

%%%%%%%%%%%%%%%%%%%%%%%%%%%%%%%%%%%%%%%%%%%%%%%%%%%%%%%%%%%%%%%%%%%%%%%
\section{The bosonized Hamiltonian}
\label{BosonizedHamiltonian}
%%%%%%%%%%%%%%%%%%%%%%%%%%%%%%%%%%%%%%%%%%%%%%%%%%%%%%%%%%%%%%%%%%%%%%%%

Bosonization is a powerful nonperturbative technique to deal with interacting fermions. In the case of two dimensional parity breaking systems, it can be implemented in terms of a dual gauge theory (see for instance ref.\ \cite{bos}  and references therein). On the other hand, the 
bosonization of a parity preserving system at finite density can be accomplished by introducing a restricted Hilbert 
space of small energy particle-hole fluctuations around the Fermi surface. In this case, the general formalism was developed in refs.\ \cite{CNF1,CNF2,Marston,Marston2}.

In this section we find a bosonic representation for the Hamiltonian system \ref{H} when a general energy dispersion relation (Eq.\ \ref{dispersion}) is considered. 
In order to establish notation and to make this paper self-contained we will first summarize the main concepts of bosonization by following refs. \cite{CNF1,CNF2}.

We define a reference state $|FS\rangle$ by applying fermionic creation operators to the vacuum state $|0\rangle$ so as to occupy all the states up to the Fermi surface,
\be
|FS\rangle=\prod_{\vk}^{\vk_F} c^{\dagger}_{\vk}|0\rangle.
\ee
We use this state to normal order all the relevant operators of the theory according to 
\be
: \hat O := \hat O -\langle FS | \hat O |FS \rangle.
\ee

The low energy behavior of the system is essentially described in terms of the particle-hole bosonic operator
\be
n_{\vq}(\vk,t)=c^{\dagger}_{\vk-\frac{\vq}{2}}(t)c_{\vk+\frac{\vq}{2}}(t),
\label{nkq}
\ee
where $\vk \approx \vk_F$ and small $\vq$ fluctuations 
are restricted to a thin shell around the Fermi surface. In fact, the approximation that defines the restricted Hilbert space of interest can be defined by the condition $q< D <\Lambda << k_F$, where $D$ is the shell thickness and $\Lambda$ is the width of the finite amount of patches used to cover the Fermi surface\cite{HaldaneFermiSee2}. These restrictions mean that the physical Hilbert space considered corresponds to a subset of excitations above $|FS\rangle$ mainly generated  by small angle scattering processes.
 
In this space the operators \ref{nkq} satisfy the following commutation relation\cite{CNF1}
\be
[n_{\vq}(\vk),n_{-\vq'}(\vk')]=\delta_{\vk,\vk'}\delta_{\vq,\vq'}\;\vq\cdot\vv_F 
\;\delta\left(\mu-\epsilon_{\vk}\right),
\label{nkqcr}
\ee
where because of the last delta function $\vk$ is constrained to lie on the Fermi surface.
For an arbitrary value of $\vq$, the operators $n_{\vq}(\vk,t)$ do not annihilate the reference state. However, 
we can define the operators, 
%\begin{widetext}
\bea
a_{\vq}(\vk_F)&=&\sum_{\vk}\frac{\Phi(\vk,\vk_F)}{\sqrt{N(\vk_F)V|\vq\cdot\vv_F|}}\times \nonumber \\
&&\left\{ n_{\vq}(\vk)\theta(\vq\cdot\vn)
+n_{-\vq}(\vk)\theta(-\vq\cdot\vn)\right\} \\
a^{\dagger}_{\vq}(\vk_F)&=&\sum_{\vk}\frac{\Phi(\vk,\vk_F)}{\sqrt{N(\vk_F)V|\vq\cdot\vv_F|}} \times \nonumber \\
&& \left\{ n_{-\vq}(\vk)\theta(\vq\cdot\vn)
+n_{+\vq}(\vk)\theta(-\vq\cdot\vn) \right\}
\label{aqkf}
\eea
%\end{widetext}
($\hat n$ is a unit vector normal to the Fermi surface at $\vk_F$). The smearing function 
$\Phi(\vk,\vk_F)$ is one, if $\vk$ belongs to the patch labeled by $\vk_F$ and zero otherwise.
In the thermodynamic limit we have 
\be
\lim_{D,\Lambda\to 0} \Phi(\vk,\vk_F)=\delta_{\vk,\vk_F},
\ee 
the local density of states $N(\vk_F)$ is given by 
\be
N(\vk_F)=\sum_{\vk} |\Phi(\vk,\vk_F)|^2 \delta(\mu-\epsilon(\vk)),
\ee
and the operators $a_{\vq}(\vk)$, $a^{\dagger}_{\vq}(\vk)$ satisfy 
\be
a_{\vq}(\vk_F)|FS\rangle=0,
\label{aFS}
\ee
\be
[a_{\vq}(\vk_F),a^{\dagger}_{\vq'}(\vk_F')]=\;\delta_{\vk_F,\vk_F'}
\left(\delta_{\vq,\vq'}+\delta_{\vq,-\vq'}\right),
\label{aconmutator}
\ee
generating the whole restricted Hilbert space of states. In this space, the fermion operator
\be
\psi(\vrr,\vk_F)=\sum_{\vq} e^{i \vq\cdot\vrr} c_{\vq}(\vk_F)
\makebox[.5in]{,} q<<\Lambda
\label{fermion}
\ee
can be written in bosonic form as\cite{CNF2}
\bea
\psi(\vrr,\vk_F)&=&\sqrt{\frac{N(\vk_F)}{\alpha}} U(\vk_F) \times \nonumber \\ 
&& e^ {\displaystyle{-\sum_{\vq}\frac{e^{-i \vq\cdot\vrr}}{N(\vk_F)V |\vq\cdot\vv_F|} 
n_{-\vq}(\vk_F)}}, 
\label{fermionnq}
\eea
or equivalently, in terms of $a_{\vq}(\vk_F)$,
\bea
\psi(\vrr,\vk_F)&=&\sqrt{\frac{N(\vk_F)}{\alpha}} U(\vk_F)\times \nonumber \\
&&e^{\displaystyle{-\sum_{\vq\cdot\vn>0}\frac{\left\{e^{-i \vq\cdot\vrr} a^{\dagger}_{\vq}(\vk_F)-e^{i \vq\cdot\vrr} a_{\vq}(\vk_F)\right\}}{\sqrt{N(\vk_F)V |\vq\cdot\vv_F|}}}},\nonumber \\
&&
\label{fermionaq}
\eea
where $\alpha$ is an ultraviolet cut-off and $U(\vk_F)$ are the ``Klein Factors'' that guarantee anticommutation relations
among operators with different $\vk_F$ (for an explicit expression of the Klein Factors see ref.\ \cite{Marston}). 

Using the fermion-boson mapping \ref{fermionaq} and the bosonic  
commutation relations \ref{aconmutator} we can get the bosonized projection of any fermionic operator onto the restricted Hilbert space of states.   
For instance, the interacting part of the Hamiltonian (Eq.\ \ref{Hint}) is simply  bosonized since it can be written in terms of $n_q(\vk_F)$, provided we  restrict the momentum space considering  only small angle scattering processes. Normal ordering the projected Eq.\ \ref{Hint} we find,
\bea
H_{int}&=& 
\frac{1}{2}\sum_{\vk_F,\vk_F',\vq} F_{\vk_F,\vk_F'}(\vq)\;|\vq\cdot\vv_F|^{1/2}|\vq\cdot\vv_F'|^{1/2}\times 
\nonumber \\
&&\left\{a^{\dagger}_{\vq}(\vk_F)a_{\vq}(\vk_F')\;\; \Theta(\vq\cdot\vk_F) \Theta(\vq\cdot\vk_F')\right.\nonumber \\
&+&\left.\;\;
a_{\vq}(\vk_F)a_{\vq}(\vk_F')\;\; \Theta(-\vq\cdot\vk_F) \Theta(\vq\cdot\vk_F')+ \mbox{h.c}
\right\},\nonumber \\
&&
\label{Hintbos}
\eea
where we have introduced the adimensional Landau function $F_{\vk_F,\vk_F'}(\vq)=N^{1/2}(\vk_F)N^{1/2}(\vk_F')f_{\vk_F,\vk_F'}(\vq)$, and $\Theta$ is the usual Heaviside function.

The restriction of the Hilbert space is usually justified by RG arguments\cite{Froehlich}. The long angle scattering coupling constants flow to zero at long distances, leaving a bosonized Hamiltonian only containing small angle scattering excitations. That is, the long angle scattering operators are perturbatively irrelevant in the Renormalization Group sense. However, they renormalize the parameters in the Hamiltonian. For this reason, the couplings $F_{\vk_F,\vk_F'}(\vq)$ in Eq.\ \ref{Hintbos} should be considered as phenomenological inputs with no trivial connection with the microscopic ones. This limitation is at the heart of the bosonization procedure. However, this technique gives very 
general and powerful results concerning phase diagrams and the universal structure of fermionic correlation functions. Of course, to make contact with microscopic models nontrivial numerical computations are necessary.

Notice that for a fixed $\vq$, the first term in Eq.\ \ref{Hintbos} represents interactions among particle-hole pairs in the same hemisphere of the Fermi surface (with respect to the direction of $\vq$), while in the second term the interaction  
mixes the two hemispheres. This second term does not contribute to the asymptotic fermionic correlation functions\cite{CNF2}, however, we will keep this term since it could become relevant in the case of nested Fermi surfaces.  

The bosonization of the free fermionic Hamiltonian is less trivial. 
Firstly, let us see what kind of terms appear in $H_0$ (Eq.\ \ref{H0}).
The tensor structure of the Hamiltonian is conveniently written in terms of a local reference frame defined by the unit vectors $\vn$ and $\vt$, normal and tangent to the Fermi surface, respectively. 
In terms of these directions it is easy to see that when considering the expansion \ref{dispersion}, the first term in $H_0$ only contains the normal field derivative
\be
\int d^nx\; \psi^{\dagger}(x,\vk_F)\; {\bf \vn\cdot\vec{\nabla}} \;\psi(x,\vk_F).
\ee 
For the second order derivatives of the dispersion relation we can write 
\be
\left.\frac{\partial^2 \epsilon}{\partial k_i\partial k_j}\right|_{\vk=\vk_F}\!\!\!\!\!\!\!=
\xi_1\;n_in_j + \xi_2\;t_it_j + \xi_3\;(n_it_j+n_jt_i),
\ee
where $\xi_1$, $\xi_2$ and $\xi_3$ are functions of $\vk_F$. A similar expression can be written for the third order rank tensor containing the third order derivatives. Therefore, the $x-\mbox{space}$ representation of the dispersive part of the Hamiltonian contains normal ($\vn\cdot\vec{\nabla}$) as well as tangential field derivatives ($\vt\cdot\vec{\nabla}$). 

In order to obtain the bosonized form of $H_0$ we consider the point-splitted product of the fermion field and its adjoint, along a general direction $\va$. Using Eq.\ \ref{fermionnq} and the Baker-Hausdorff 
formula
\be
e^{\hat A}e^{\hat B}= : e^{\hat A+\hat B} : e^{\langle\hat A\hat B +\frac{1}{2}(\hat A^2+\hat B^2)\rangle},
\ee
we obtain,
\bea
\lefteqn{
: \psi^{\dagger}\left(\vrr-\frac{1}{2}\va,\vk_F\right)\psi\left(\vrr+\frac{1}{2}\va,\vk_F\right):= 
e^{\displaystyle{G(\va,\vk_F)}}\times}\nonumber \\
&& \nonumber \\
&& : e^{-2 i \displaystyle{\sum_{\vq} \frac{
e^{i\vq\cdot\vrr}}{N(\vk_F) V \vq\cdot\vv_F}}
\sin\left(\frac{\vq\cdot\va}{2}n_{\vq}(\vk_F)\right)}, \!\!\!\!\!\!\!\!\!\!\!\!:
\nonumber \\
&&
\label{point-splitting}
\eea
where
\be
G(\va,\vk_F)=\sum_{\vq\cdot\vn>0} \frac{\left(e^{i\vq\cdot\va}-1\right)}{N(\vk_F) V |\vq\cdot\vv_F|}.
\label{G}
\ee
Let us first consider a direction $\va$ locally tangent to the Fermi surface ($\va= \epsilon \vt$). 
In this case, when summing in Eq.\ref{G} over the normal ($q_N=\vq\cdot\vn$) and tangent ($q_T=\vq\cdot\vt$) components we get, 
\be
G(\epsilon,\vk_F)=\frac{1}{N(\vk_F) V v_F}\sum_{\vq_T} \left(e^{i\epsilon \vq_T}-1\right)
\sum_{q_N>0}\frac{1}{|\vq_N|},
\ee
and considering $\epsilon\vq_T<< 1$, we obtain
\be
G(\epsilon,\vk_F)=-\frac{(\epsilon\Lambda)^2}{2 v_F} 
\int_{0}^{D} dq_N\;\frac{1}{|\vq_N|}
\ee
(for an anisotropic Fermi surface $v_F=|\vv_F|$ is $\vk_F$ dependent).
Note that the remaining integral in $q_N$  has a logarithmic infrared divergence,
$G(\epsilon,\vk_F)\propto -\ln(V)$, and since the normal order in Eq.\ref{point-splitting} is a regular function we conclude that in the thermodynamic limit: 
\be
: \psi^{\dagger}\left(\vrr-\frac{1}{2}\epsilon\vt,\vk_F\right)\psi\left(\vrr+\frac{1}{2}\epsilon\vt,\vk_F\right):= 
0.
\ee
This result implies that the tangent derivatives do not contribute to the projected bosonized Hamiltonian. In this regard, we recall that the bosonic particle-hole excitations tangent to the Fermi surface do not either contribute to the asymptotic form of the correlation function\cite{CNF2}, however they do contribute to the density of states and they are crucial to obtain the correct specific heat and other thermodynamic properties\cite{CNF1}. 

We will now concentrate on the bosonization of fermionic terms containing normal derivatives. In this case we can 
generalize a calculation proposed by Haldane in one spatial dimension\cite{Haldane} to the case of arbitrary dimensions.
Let us consider the integral, 
\bea
A&=&\int d\vrr\; 
:\psi^{\dagger}\left(\vrr-\frac{1}{2}\epsilon\;\vn,\vk_F\right)\psi\left(\vrr+\frac{1}{2}\epsilon\;\vn,\vk_F\right):
\nonumber \\
&& \label{A}
\eea
Introducing Eq.\  \ref{fermion} into Eq.\ \ref{A} and expanding in powers of $\epsilon$ we find on the one hand, 
\be
A=\sum_n \frac{(-i)^n}{n!}\epsilon^n\sum_{\vq}\left(\vq\cdot\vn\right)^n :c^{\dagger}_{\vq}(\vk_F) c_{\vq}(\vk_F):.
\label{AF}
\ee
On the other hand, choosing $\va=\epsilon\;\vn$ in Eq.\ \ref{point-splitting},
replacing into Eq.\ \ref{A} and expanding in powers of $\epsilon$ we find the bosonic version of $A$. Thus, comparing these two expressions order by order in $\epsilon$, we find the bosonized projected Hamiltonian. When expanding the dispersion relation up to third order in the derivatives we get
\be
H_0=\sum_{\vk_F} h_1(\vk_F)+ h_2(\vk_F)+ h_3(\vk_F),
\label{H0B}
\ee
with
\begin{widetext}
\bea
h_1&=&\frac{v_F}{2}\sum_{\vq\cdot\vn>0} |\vq\cdot\hat{n}|\;\; a^{\dagger}_{\vq}(\vk_F)a_{\vq}(\vk_F)
\label{h1}
\\
h_2&=&\frac{\beta}{2 (N(\vk_F)V v_F)^{1/2}}
\sum_{\vq_i}
|\vq_1\cdot\hat{n}|^{1/2}|\vq_2\cdot\hat{n}|^{1/2}|\vq_3\cdot\hat{n}|^{1/2}\;\;\delta(\vq_1+\vq_2+\vq_3)
\times \nonumber \\
&&\left\{ a^{\dagger}_{\vq_1}(\vk_F)a^{\dagger}_{\vq_2}(\vk_F)a_{\vq_3}(\vk_F)\;
\Theta(-\vq_1\cdot\hat{n})\Theta(-\vq_2\cdot\hat{n})\Theta(\vq_3\cdot\hat{n}) + \mbox{h.\ c.\ }\right\}
\nonumber \\
&& \label{h2}
\\
h_3&=&\frac{\gamma}{4!\; N(\vk_F)V v_F}
\sum_{\vq_i}
|\vq_1\cdot\hat{n}|^{1/2}|\vq_2\cdot\hat{n}|^{1/2}|\vq_3\cdot\hat{n}|^{1/2}
|\vq_4\cdot\hat{n}|^{1/2}\;\;\delta(\vq_1+\vq_2+\vq_3+\vq_4)
\times \nonumber \\
&&\left\{4\;a^{\dagger}_{\vq_1}(\vk_F)a_{\vq_2}(\vk_F)a_{\vq_3}(\vk_F)a_{\vq_4}(\vk_F)
\Theta(-\vq_1\cdot\hat{n})\Theta(\vq_2\cdot\hat{n})\Theta(\vq_3\cdot\hat{n})
\Theta(\vq_4\cdot\hat{n})\right.\nonumber \\
&&+\left.
3\; a^{\dagger}_{\vq_1}(\vk_F)a^{\dagger}_{\vq_2}(\vk_F)a_{\vq_3}(\vk_F)a_{\vq_4}(\vk_F)
\;\Theta(-\vq_1\cdot\hat{n})\Theta(-\vq_2\cdot\hat{n})\Theta(\vq_3\cdot\hat{n})
\Theta(\vq_4\cdot\hat{n})
+ \mbox{h.\ c.\ } \right\}, 
\label{h3}
\eea
\end{widetext}
where 
\bea
\beta&=& \left.\frac{\partial^2 \epsilon (\vk)}{\partial k_j\partial k_i}\vn_i\vn_j\right|_{\vk=\vk_F}
 \label{beta}\\
\gamma&=&\left.\frac{\partial^3 \epsilon (\vk)}{\partial k_l\partial k_j\partial k_i}\vn_i\vn_j\vn_l\right|_{\vk=\vk_F}.
\label{gamma}
\eea
The coefficient $\beta$ is related to particle-hole asymmetry. In fact, in a system invariant under charge conjugation, $\epsilon(\vk_F+\vq)-\mu$ is odd under the transformation $\vq \to -\vq$. In this case, the even derivatives of $\epsilon$
at the Fermi level ($\vk =\vk_F$) vanish.  For stability reasons we will consider $\gamma>0$. If $\gamma$ happens to be negative, then we should continue 
expanding the fermion dispersion relation until a well defined result be achieved.  

Eqs.\ \ref{H0B} to \ref{h3} display the main results of this section. The $h_1$-term corresponds to the bosonized free fermion Hamiltonian when a linear dispersion relation is considered and coincides with that one computed in refs. \cite{CNF1,Marston}. Then, the final bosonized Hamiltonian contains a quadratic part 
($\sum_{k_F} h_1+H_{\rm int}$) plus a nonquadratic ($h_2+h_3$)-term which is related to dispersive effects on free fermions.
In order to calculate any observable the dispersive part should be treated by means of perturbation theory. In the Fermi liquid regime the nonquadratic terms are irrelevant, although interesting effects were studied in the context of Landau theory\cite{CN}.

Here, we are interested in understanding the role played by the presence of dispersion effects when the interacting system is otherwise unstable. The next two sections are devoted to study this issue. 
%%%%%%%%%%%%%%%%%%%%%%%%%%%%%%%%%%%%%%%%%%%%%%%%%%%%%%%%%%%%%%%%%%%%%%%%%%%%%%%%%
\section{Dynamics of the Fermi surface}
\label{Dynamics}
%%%%%%%%%%%%%%%%%%%%%%%%%%%%%%%%%%%%%%%%%%%%%%%%%%%%%%%%%%%%%%%%%%%%%%%%%%%%%%%%%%

As shown in ref.\  \cite{CNF1}, the Fermi surface deformations can be associated to collective particle-hole excitations described by coherent sates in the bosonized theory. In the stable case, contact is made with the Landau theory for Fermi liquids. As we will see, this procedure applies equally well to the general case where dispersion effects must be taken into account.

Following ref.\ \cite{CNF1}, we define a many-body state which is a direct product of coherent states parametrized by a complex field $\phi_{\vq}(\vk)$:
\be
|\phi\rangle=U(\phi)|FS\rangle, 
\ee
where 
\be
U(\phi)=e^{-\Gamma(\phi)}
\ee
and
\bea
\Gamma(\phi)&=&\!\!\!\!\!\!\!\!\!\!\sum_{\vk_F,\vq\cdot\vk_F>0}\left(\frac{1}{N(\vk_F) V \vq\cdot\vf} \right)^{1/2}
\times\nonumber \\
&&\left\{ \phi_{\vq}(\vk_F) a^{\dagger}_{\vq}(\vk_F)-\phi^*_{\vq}(\vk_F) a_{\vq}(\vk_F)\right\}.
\eea
The ``deformation'' fields satisfy $\phi^*_{\vq}(\vk_F)=\phi_{-\vq}(\vk_F)$, implying $\Gamma^*(\phi)=-\Gamma(\phi)$ and 
$U^{\dagger}(\phi)=U^{-1}(\phi)$.

The system's partition function can be written in terms of this overcomplete coherent state basis by means of the path integral  
\be
Z=\int {\cal D}\phi {\cal D}\phi^*\; e^{i S(\phi,\phi^*)},
\ee 
where $S(\phi,\phi^*)=\int dt~{\cal L}(\phi,\phi^*)$ and 
\bea
{\cal L}(\phi,\phi^*)&=&\sum_{\vk_F,\vq}\frac{i}{N(\vk_F) V |\vq\cdot\vf|}
\left(\phi^*_{\vq}(\vk_F,t)
\frac{\partial \phi_{\vq}(\vk_F,t)}{\partial t}\right)- \nonumber \\
&-& \langle \phi|H|\phi\rangle.
\label{lagrangianaq}
\eea

The evaluation of $\langle \phi|H|\phi\rangle$ is straightforward, since the bosonized Hamiltonian (Eq.\ \ref{H0B}) is normal ordered and 
the coherent states are eigenvalues of the destruction operator $a_{\vq}(\vk_F)$:  
\bea
a_{\vq}(\vk_F)|\phi\rangle&=&\frac{\left(\phi_{\vq}(\vk_F) \Theta(\vq\cdot\vk_F)+\phi^*_{\vq}(\vk_F) \Theta(-\vq\cdot\vk_F)\right)}{\sqrt{N(\vk_F) V |\vq\cdot\vf|}}
 \;|\phi\rangle \nonumber \\
\label{aphi}
\eea

Using Eqs. \ref{Hintbos}, \ref{H0B} and \ref{aphi}, we find the following Lagrangian (after the field redefinition 
$\phi\to (V N(\vk_F)v_F)^{1/2} \phi $):
\bea
\lefteqn{
{\cal L}= \sum_{\vq \vk}\left(\frac{i}{\vq\cdot\vn} \right)  
\phi^*_{\vq}(\vk,t)
\frac{\partial \phi_{\vq}(\vk,t)}{\partial t}} \nonumber \\
+&\frac{1}{2}&\!\!\!\!\sum_{\vq \vk\vk'}\!\!\phi^*_{\vq}(\vk)\left\{v_F\delta_{\vk,\vk'}+F_{\vk,\vk'}(\vq)\sqrt{v_F
v_F'} \right\}\phi_{\vq}(\vk')
\nonumber \\
+&&\Re  \left( \frac{\beta}{3!}\sum_{\vk\vq_i}
\phi_{\vq_1}(\vk)\phi_{\vq_2}(\vk)\phi_{\vq_3}(\vk) \delta(\vq_1+\vq_2+\vq_3)+\right.
\nonumber \\
+&&\frac{\gamma}{4!}\left. \sum_{\vk\vq_i}
\phi_{\vq_1}(\vk)\phi_{\vq_2}(\vk)\phi_{\vq_3}(\vk)\phi_{\vq_4}(\vk) \delta(\vq_1+\dots+\vq_4)\right),
\nonumber \\
&&
\label{lagrangianaqfinal}
\eea
where $\vk$ and $\vk'$ lie on the Fermi surface and we have absorbed powers of $N(0)V v_F$\cite{density} into the definition of $\beta$ and $\gamma$.

The Lagrangian in Eq.\ \ref{lagrangianaqfinal} controls the low energy dynamics of the Fermi surface.
Note that this dynamics is extremely nonlocal, due to the factor $1/|\vq\cdot\vn|$ in the kinetic term.
In the region where the quadratic term is stable (positive definite), the first two lines represent the usual Landau theory of Fermi liquids. This is a fixed point in the RG sense, that is, the nonquadratic dispersive terms do not modify the asymptotic correlation functions. However, if interactions are such that the quadratic term is unstable, the nonquadratic terms become relevant, stabilizing the theory in another fixed point. 

A word of caution is necessary to understand the validity of Eq. \ref{lagrangianaqfinal}. 
According to the Renormalization Group theory, not only the couplings $F_{\vk,\vk'}(\vq)$ but also the parameters $\beta$ and $\gamma$ will be renormalized by irrelevant operators.
Therefore, although we showed how the band curvature generates the nonquadratic terms in Eq.\ \ref{lagrangianaqfinal} (see Eqs.\ \ref{beta} and \ref{gamma}), the actual calculation of these parameters for a given band structure is not at all trivial. 

Also, in the present paper we are interested in understanding how the band curvature can drive the system to a new rotational symmetry breaking ground state. For this reason, in Eq.\ \ref{lagrangianaqfinal}, we have only considered  small angle scattering processes with small momentum transfer, disregarding any irrelevant term coming from the interaction Hamiltonian. 
On the other hand, it is well known that some interaction channels, although irrelevant, could give raised to different instabilities. For instance, the Kohn-Luttinger instability\cite{Kohn-Luttinger} comes from the competition of forward scattering with a BCS channel, even for repulsive interactions. Recently, a similar dynamical effect in $2d$ Fermi liquids was reported in ref.\ \cite{Kohn-Luttinger-Sarma}. 
The interplay among the anisotropic phases studied here and other interaction channels is a very interesting issue, and the multidimensional bosonization technique described in this paper seems to be a promising tool to face it. 

With this comments in mind, we can study for instance the static deformations of the Fermi surface described by the 
system's free energy, which can be computed as the action 
per unit time, when setting to zero the kinetic term in Eq.\ \ref{lagrangianaqfinal}. Introducing the Fourier transformed field 
\be
\varphi(x,\vk)=\int \frac{d^dq}{(2\pi)^d}\; \phi_{\vq}(\vk)\;\; e^{-i \vq\cdot\vec{x}},
\label{FT}
\ee
at each point of the Fermi surface, the expression for the free energy is simplified to
\bea
F&=&\frac{1}{2}\int dxdx' \sum_{\vk\vk'} \varphi(x,k)
M(\vk-\vk',x-x')\varphi(x',k')\nonumber \\
&+&\sum_{\vk}\int dx \left\{\frac{\beta}{3!}\varphi(x,k)^3+\frac{\gamma}{4!}\varphi(x,k)^4\right\},
\label{FreeEnergy}
\eea
where 
\bea
\lefteqn{
M(\vk-\vk',x-x')=}\nonumber \\
&&\left\{v_F\delta_{\vk,\vk'}\delta(x-x')+F_{\vk,\vk'}(x-x')\sqrt{v_F
v_F'} \right\}. 
\nonumber \\
\eea
In the next section, using the framework provided by bosonization, the problem of spontaneously broken rotational 
invariance is explored.

%%%%%%%%%%%%%%%%%%%%%%%%%%%%%%%%%%%%%%%%%%%%%%%%%%%%%%%%%%%%%%%%%%%%%%%
\section{The isotropic-nematic-hexatic quantum phase transitions}
\label{isotropic-nematic-hexatic}
%%%%%%%%%%%%%%%%%%%%%%%%%%%%%%%%%%%%%%%%%%%%%%%%%%%%%%%%%%%%%%%%%%%%%%%

%%%%%%%%%%%%%%%%%%%%%%%%%%%%%%%%%%%%%%%%%%%%%%%%%%%%%%%%%%%%%%%%%%%%%%%%%%%
\begin{figure}
\begin{center}
\leavevmode
\vspace{1cm}
\noindent
\hspace{1.0 in}
\epsfxsize=8 cm
\epsfysize=8 cm
\epsfbox{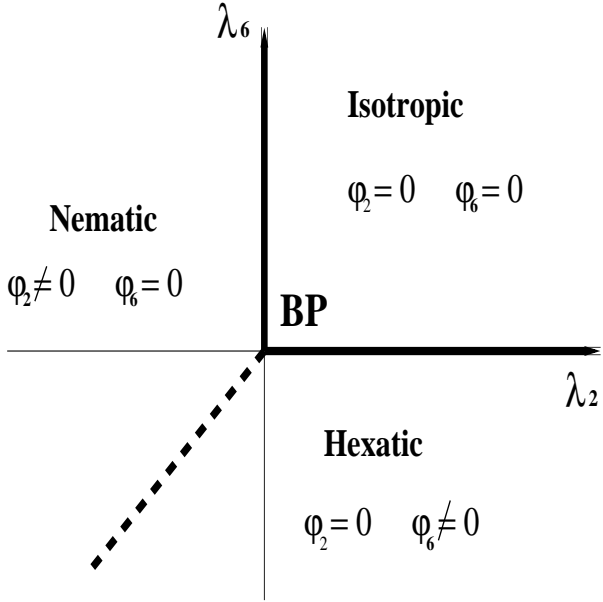}
\vspace{1cm}
\end{center}
\caption
{Phase diagram of an electron gas with nematic and hexatic instabilities. In this case $|F_n|<<1$, for $n\ne 2, 6$. 
The bold and dot lines represent second and first order transitions, respectively. The origin $BP$ is a bicritical point}
\label{bp}
\end{figure}
%%%%%%%%%%%%%%%%%%%%%%%%%%%%%%%%%%%%%%%%%%%%%%%%%%%%%%%%%%%%%%%%%%%%%%%%%%%%%%%%

The order parameter of the two dimensional nematic phase is a second-rank traceless antisymmetric tensor. It is odd under $\pi/2$ spatial rotations, and the two independent components can be cast in the form of a ``headless'' two dimensional vector called ``director''. The director is invariant under a $\pi$ rotation, characterizing in this way the nematic state\cite{deGennes}. On the other hand, an hexatic phase is characterized by an order parameter that is invariant under $\pi/3$ rotations. These states are of great interest as they are candidates to present non-Fermi liquid behavior in two dimensions\cite{NematicKFV}.

By means of the bosonization approach, we will study them as particular instabilities of an isotropic Fermi surface.
The important role played by the fermionic interactions as well as the dispersion effects for the establishment of these 
phases will become explicit.

Let us begin by  considering an isotropic Fermi surface in two dimensions and a short ranged interaction. In this case, the Landau function only depends on the angle between the two Fermi vectors $\vk$ and $\vk'$
\be
F_{\vk,\vk'}(x-x')=F_{\theta-\theta'}\;\delta(x-x').
\ee
With these considerations, the ground state of Eq.\ \ref{FreeEnergy} becomes homogeneous, $\langle\varphi(x,k)\rangle=\varphi(\theta)$, and the free energy is simplified to
\bea
F&=&\frac{v_F}{2}\int d\theta d\theta'\;\; \varphi(\theta)
\left\{\delta_{\theta-\theta'}+ F(\theta-\theta')\right\}\varphi(\theta')
\nonumber \\
&+&\int d\theta \left\{\frac{\beta}{3!}\varphi(\theta)^3+\frac{\gamma}{4!}\varphi(\theta)^4\right\}.
\label{Ftheta}
\eea
Now, introducing the representation
\bea
\varphi(\theta)&=&\sum_n \varphi_n e^{i n \theta} \\
F(\theta)&=&\sum_n F_n e^{i n \theta}
\eea
($\varphi_n^*=\varphi_{-n}$) and considering a particle-hole symmetric system ($\beta=0$) we obtain,
\bea
F=&\frac{v_F}{2}&\!\!\!\!\sum_n \left(1+ F_n \right)\varphi_n\varphi_n 
+\frac{\gamma}{4!}\!\sum_{nmlp}\!\!\varphi_n\varphi_m\varphi_l\varphi_p\delta_{n+m+l+p}.
\nonumber \\
&&
\label{Fn}
\eea

%%%%%%%%%%%%%%%%%%%%%%%%%%%%%%%%%%%%%%%%%%%%%%%%%%%%%%%%%%%%%%%%%%%%%%%%%%%
\begin{figure}
\begin{center}
%\leavevmode
\vspace{1cm}
\noindent
\hspace{1.0 in}
\epsfxsize=8 cm
\epsfysize=8 cm
\epsfbox{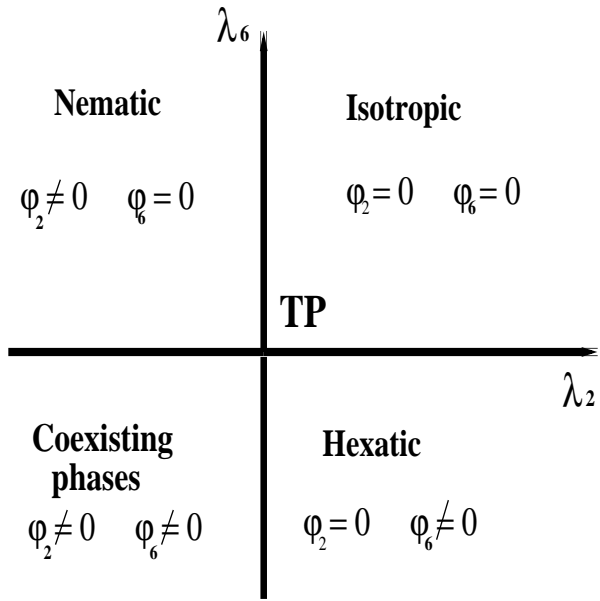}
\vspace{1cm}
\end{center}
\caption{Phase diagram of an electron gas with nematic and hexatic instabilities. In this case, the Landau parameters 
are not restricted to small values and $\mu_R>\gamma_R^2/16$. The bold lines represent second order phase 
transitions and $TB$ is a tetracritical point}
\label{tp}
\end{figure}
%%%%%%%%%%%%%%%%%%%%%%%%%%%%%%%%%%%%%%%%%%%%%%%%%%%%%%%%%%%%%%%%%%%%%%%%%%%%%%%%

In the case where $F_n\ge -1$, it is clear that the only minimum of Eq.\ \ref{Fn} corresponds to $\varphi_n=0$, for all $n$. This is the usual criterium for the Fermi liquid stability\cite{CNF1,HaldaneFermiSee1,Pomeranchuk,Shankar}. However, if for some channel $n$ it happens that $F_n< -1$, then it is possible to develop quantum phase transitions to anisotropic states. 
Suppose for instance that we have a ground state characterized by $\varphi_2\ne 0$, and 
$\varphi_n= 0$ for all $n\ne 2$. In this case, the deformation field would have the form 
\be
\varphi(\theta)=4\varphi_2\;\;\left\{2 \cos^2(\theta)-1\right\}, 
\ee
characterizing a two dimensional nematic order\cite{deGennes}. This would correspond to a transition where, because of the interactions, the otherwise circular shape of the Fermi surface is deformed into an ellipse.  

In general, as we mentioned above, it is interesting to study instabilities in the nematic/hexatic sector. They are 
expected to occur when $F_2$ and $F_6$ are large and negative\cite{NematicKFV}. This case can be treated by integrating over all  
the stable modes $\varphi_n$, $n\ne$ $2$, $6$. As a result of this integration all the relevant terms compatible with the symmetry will be generated, obtaining the following effective free energy for the modes $\phi_2$ and $\phi_6$, 
\bea
F=\lambda_2 \varphi_2^2 +\lambda_6 \varphi_6^2+ 
\frac{\gamma_R}{4}\left\{\varphi_2^4 
 + \varphi_6^4\right\}+ \mu_R\varphi_2^2\varphi_6^2,
\eea
which is written in terms of renormalized interaction constants ($\lambda_2= \bar{v}_F(1+ F_2)$, 
$\lambda_6=\bar{v}_F(1+F_6$). This free energy presents the interesting possibility of quantum phase transitions\cite{condensedmatter} among isotropic ($\varphi_2=\varphi_6=0$), nematic ($\varphi_2\ne 0~,~\varphi_6=0$) and hexatic states ($\varphi_2= 0~,~\varphi_6\ne0$). There is also the possibility of coexisting nematic-hexatic phases 
($\varphi_2\ne 0~,~\varphi_6\ne 0$).

It is important to note that while the phase transitions are triggered by the values of $F_2$ and $F_6$, 
the structure of the phase diagram depends on the other Landau parameters $F_n$  ($n\ne 2$, $6$) hidden in $\gamma_R$ and $\mu_R$.
For instance, if $|F_n|<< 1$ (for all $n\ne 2,6$), then $\mu_R\approx \gamma_R/2$, and the qualitative phase diagram 
is that shown in figure \ref{bp}. This diagram presents a bicritical point at the origin of the $\lambda_1-\lambda_2$ plane. The bold lines represent second order phase transitions between the isotropic/nematic and the isotropic/hexatic phases. Moreover, the interphase nematic/hexatic (doted line in Fig.\ \ref{bp}) corresponds to a first order phase transition.  

However, if $F_n \approx 1$ there is the possibility of having $\gamma_R^2/16 < \mu_R$. In this case the phase diagram changes qualitatively (see Fig.\ \ref{tp}). There is a tetracritical point with four second order phase transitions. For $\lambda_1<0, \lambda_2<0$ there is a region of coexisting phases which is absent in the preceding case. 

In order to study the electronic properties of these phases we need to consider the dynamics associated with the Lagrangian Eq.\ \ref{lagrangianaqfinal} and to evaluate quantum fluctuations around the saddle points found in these sections. This discussion will be presented elsewhere.

%%%%%%%%%%%%%%%%%%%%%%%%%%%%%%%%%%%%%%%%%%%%%%%%%%%%%%%%%%%%%%%%%%%%%
\section{Discussion and conclusions}
\label{Conclusions}
%%%%%%%%%%%%%%%%%%%%%%%%%%%%%%%%%%%%%%%%%%%%%%%%%%%%%%%%%%%%%%%%%%%%%

In this work we have constructed the bosonized action for a general fermionic system having a smooth Fermi surface and a nonlinear energy dispersion relation in higher dimensions.
We have shown that the effect of the nonlinear terms in the energy dispersion is that of producing interactions in the bosonized theory.  To the best of our knowledge this is the first explicit generalization to higher dimensions of the 
well known analogous result in one dimension\cite{Haldane}.

The Hamiltonian \ref{H0B}, the corresponding action in the coherent state basis \ref{lagrangianaqfinal} and the free energy in Eq.\  \ref{FreeEnergy} are the main results of this paper. 

~From a physical point of view, dispersion effects are irrelevant when the system is in the normal Fermi liquid regime, however, they are essential when due to some fermion interactions an instability of the Fermi surface occurs, driving the electronic system outside this regime. In this later case, the induced nonquadratic terms in the bosonized free energy \ref{FreeEnergy} will stabilize the electronic system in a new ground state. 

In particular we have concentrated in two quantum liquid crystal states: nematic and hexatic.
The corresponding phase transitions are triggered by negative values of the Landau parameters $F_2$ and $F_6$, associated with the fermion interaction\cite{NematicKFV}. However, the qualitative structure of the phase diagrams (Figs. \ref{bp} and \ref{tp}) depends on the relative values of all the other ``stable'' Landau parameters. When they are small, 
the phase diagram has a tricritical point where two second order phase transitions (isotropic/nematic, isotropic/hexatic) and a first order one (nematic/hexatic) meet together. In the opposite case, there is the possibility of having a coexisting nematic-hexatic phase with  a tetracritical point. 

These results were obtained by means of a mean field saddle-point calculation on the bosonized action. It is clear that 
some of the global features (specially the first order phase transition in Fig.\ \ref{bp}) could be modified by quantum fluctuations. 

In particular, it would be very interesting to investigate the dynamical electronic properties of these phases. In this regard, it was shown\cite{NematicKFV} that the Goldstone modes of the nematic phase are damped (except for certain Fermi points dictated by symmetry considerations) while the fermion correlation function shows non-Fermi Liquid behavior.  
In order to address these points in the nonperturbative bosonization framework it is necessary to introduce quantum fluctuations around each saddle-point and employ the Lagrangian \ref{lagrangianaqfinal} and the fermionic operator \ref{fermionaq} to compute fermion correlators. We hope to present results on this issue soon.

\section{Acknowledgments}

We are very grateful to Eduardo Fradkin and A.\ H.\ Castro Neto for useful comments. 

The Conselho Nacional de Desenvolvimento Cient\'{\i}fico e Tecnol\' ogico CNPq-Brazil, the Funda\c c\~ ao de Amparo \`a
Pesquisa do Estado do Rio de Janeiro (Faperj) and the SR2-UERJ are acknowledged for the financial support.

%%%%%%%%%%%%%%%%%%%%%%%%%%%%%%%%%%%%%%%%%%%%%%%%%%%%%%%%%%%%


\begin{thebibliography}{99}
\bibitem{density} For smooth Fermi surface (free of Van Hove singularities) we can substitute the local density of states 
$N_{\vk_F}$ by the total density of states $N(0)$. The quotient between these two quantities is just the solid angle on the Fermi surface $S_d=\int d\Omega=\frac{N(0)}{N_{\vk_F}}$ (see ref.\ \cite{CNF2}, pag.\ 4087).
\end{thebibliography}
\end{document}